\documentclass[]{article}
\usepackage{smc2020}
\usepackage{times}
\usepackage{ifpdf}
\usepackage[english]{babel}
\usepackage{cite}
\usepackage{balance}

\def\papertitle{ASMD: AN AUTOMATIC FRAMEWORK FOR COMPILING MULTIMODAL DATASETS WITH AUDIO AND SCORES}
\def\firstauthor{Federico Simonetta}
\def\secondauthor{Stavros Ntalampiras}
\def\thirdauthor{Federico Avanzini}

\newif\ifpdf
\ifx\pdfoutput\relax
\else
   \ifcase\pdfoutput
      \pdffalse
   \else
      \pdftrue
\fi

\ifpdf % compiling with pdflatex
  \usepackage[pdftex,
    pdftitle={\papertitle},
    pdfauthor={\firstauthor, \secondauthor, \thirdauthor},
    bookmarksnumbered, % use section numbers with bookmarks
    pdfstartview=XYZ % start with zoom=100% instead of full screen; 
   ]{hyperref}
  \usepackage[pdftex]{graphicx}
  \DeclareGraphicsExtensions{.pdf,.jpeg,.png}

  \usepackage[figure,table]{hypcap}

\else % compiling with latex
  \usepackage[dvips,
    bookmarksnumbered, % use section numbers with bookmarks
    pdfstartview=XYZ % start with zoom=100% instead of full screen
  ]{hyperref}  % hyperrefs are active in the pdf file after conversion

  \usepackage[dvips]{epsfig,graphicx}
  \usepackage[figure,table]{hypcap}
\fi

\hypersetup{
    colorlinks,%
    citecolor=black,%
    filecolor=black,%
    linkcolor=black,%
    urlcolor=black
}

\title{\papertitle}

\oneauthor
   {{\firstauthor}\qquad{\secondauthor}\qquad{\thirdauthor}} {University of Milan \\ Department of Computer Science \\
   LIM - Music Informatics Laboratory \\
         {\tt [name].[surname]@unimi.it}}

\usepackage[frozencache,cachedir=.]{minted}
\usepackage{stfloats}
\usepackage{paralist}
\begin{document}
\capstartfalse
\maketitle
\capstarttrue
%
%\begin{multicols}{2}
\begin{abstract}
    This paper describes an open-source Python framework for handling datasets for music processing tasks, built with the aim of improving the reproducibility of research projects in music computing and assessing the generalization abilities of machine learning models. The framework enables the automatic download and installation of several commonly used datasets for multimodal music processing. Specifically, we provide a Python API to access the datasets through Boolean set operations based on particular attributes, such as intersections and unions of composers, instruments, and so on. The framework is designed to ease the inclusion of new datasets and the respective ground-truth annotations so that one can build, convert, and extend one's own collection as well as distribute it by means of a compliant format to take advantage of the API. All code and ground-truth are released under suitable open licenses.
\end{abstract}

\section{Introduction}\label{sec:introduction}
A recent trend in computer science is the adoption of multimodal strategies for increasing the effectiveness of algorithmic solutions in several domains\cite{baltrusaitis2018multimodal,8816973,Ntalampiras2012,atrey2010multimodal,minsky1991logical}. This comes as a natural consequence of the a) ever-increasing availability of computational resources, which are now able to deal with big data, and b) popularity of machine learning algorithms, the performance of which is boosted as more data (including multimodal) becomes available.
As a result, machine learning technologies are now employed in novel and unexplored ways. 
%thanks to the modern computers and devices which are becoming more and more affordable.

In the context of music information processing, several tasks still pose unsolved challenges to the research community, and multimodal approaches could provide a promising path. The fields of \textit{multimodal music processing} and \textit{multimodal music representation} have already been investigated in previous works\cite{simonetta201901multimodal, ludovico2019adoption}.

Two issues that are more and more debated in several research fields 
%, including multimodal approaches methods, 
are the ability to \textit{reproduce} published results\cite{baker2016scientists,hutson2018artificial} and to \textit{generalize} the resulting models\cite{abumostafa2012learning}.

Reproducibility is associated with the differences occurring in various implementations of the same method. As an example, one issue is related to the different data formats used in music and in the available datasets, which might cause troubles in the translation between representation formats and, consequently, in the reproducibility of research.

The generalization problem instead is due, among other factors, to the need of large and well-annotated datasets for training effective models. In particular, the whole field of music information processing has only a limited number of large datasets which could be much more useful if they could be merged together. Music itself, moreover, is particularly affected by the difficulty of creating accurate annotations to evaluate and train models, often hindering the collection of large datasets and causing a low generalization ability. 

With these three keywords in mind (\textit{multimodal}, \textit{reproducibility} and \textit{generalization}), we have built ASMD to help researchers in the standardization of music ground-truth annotations and in the distribution of datasets. ASMD is the acronym for Audio-Score Meta-Dataset and provides a framework for describing, converting, and accessing a single dataset which includes various datasets -- hence the expression \textit{Meta-Dataset}; it was born as a side-project of a research about audio-to-score alignment and, consequently, it contains audio recordings and music scores for all the data included in the official release -- hence the \textit{Audio-Score} part. However, we have endeavoured to make ASMD able to include any contribute from anyone. ASMD is available under free licenses.\footnote{Code is available at \url{https://framagit.org/sapo/asmd/}, documentation is available at \url{https://asmd.readthedocs.io/}}

A similar effort is held by \texttt{mirdata}~\cite{bittner2019mirdata}, a Python package for downloading and using common MIR datasets. However, our work is more focused on multimodality and tries to keep the entire framework easily extensible and modular.

In the following sections, we describe a) the design principles, b) the implementation details, c) a few use cases and, d) possible future works.

\section{Design Principles and Specifications}\label{sec:design}

In this section we present the principles which guided the design of the framework. Throughout this paper, we are going to use the word \textit{annotation} to refer to any music-related data complementing an audio recording. For instance, common types of annotations are music notes, f0 for each audio frame, beat position, separated audio tracks, etc.

\subsection{Generalization}\label{sec:generalization}
With \textit{generalization}, we mean the ability of including different datasets which are distributed with various formats and annotation types in the model generation process. This is an important issue especially during the conversion procedure: since we aimed at distributing a conversion script to recreate annotations from scratch for the sake of \textit{reproducibility}, we need to be able to handle all various storage systems -- e.g. file name patterns, directory structures, etc. -- and file formats -- e.g. midi, csv, musicxml, ad-hoc formats, etc.

Also, our ground-truth format should be generic enough to represent all the information contained in the available ground-truths and, at the same time, it should permit to handle datasets with different ground-truth types -- i.e. one dataset could provide \textit{aligned notes} and \textit{f0}, while another one could provide \textit{aligned notes} and \textit{beat-tracking}, and they should be completely accessible.

\subsection{Modularity}\label{sec:modularity}
\textit{Modularity} refers to the re-use of parts of the framework in different contexts. Modularity is important during both addition of new datasets and usage of the API. To ease the conversion between ground-truth file formats, the user should be able to re-use existing utilities to include additional datasets. Moreover, the user should be allowed to use only some parts of the datasets and the corresponding annotations.

\subsection{Extensibility}\label{sec:extensibility}
The purpose of the framework is to create a tool to help the stan\-dard\-iza\-tion of mu\-sic in\-for\-ma\-tion pro\-cess\-ing re\-search. Consequently, we aimed for a framework that is completely open to new additions: it should be easy for the user to add new datasets without editing sources from the framework itself. Also, it should be easy to convert from existing formats in order to take advantage of the API and to be able to merge existing datasets. Finally, the framework should provide a usable format to add new annotations so that new datasets can be natively created with the incorporated tools.

\subsection{Set operability}\label{sec:setOperability}
Since the framework aims at merging multiple datasets, we wanted to add the ability to perform set operations over datasets. As an example, within the context of \textit{automatic music transcription} research, several large datasets exist consisting of piano music\cite{hawthorne2019enabling,muller2011saarland,goebl1999vienna}, but only few and considerably smaller are available for other instruments\cite{thickstun2018invariances,miron2016scoreinformed,brum2018traditional,duan2011soundprism,fritsch2012trios}. Consequently, a useful feature of the framework would be the ability to select only some songs from multiple data\-sets based on particular attributes, such as the instrument involved, the number of instruments, the composer or the type of ground-truth available for that song.

\subsection{Copyrights}\label{sec:copyrights}
A common issue with distributing music recordings and annotations are copyrights. Today, most of the datasets typically used for music information processing are released under Creative Commons Licenses, but there are many exceptions of datasets released under closed terms\cite{simonetta2018symbolic,miron2016scoreinformed} or not released at all because of copyright restrictions\cite{simonetta2019convolutional}. To overcome this problem, we wanted all datasets to be downloadable from their official sources, in order to avoid any form of redistribution. Nonetheless, all the annotations that we produced were redistributable under Creative Commons License.

\subsection{Audio-score oriented}\label{sec:alignmentprinciple}
Besides the effort to produce a general framework for music processing experiments, this project was born as a utility during conducting research addressing the audio-to-score problem. The underlying idea is that we have various scores and large amounts of audio available to end-users, thus trained models could easily take advantage of such multimodality (i.e. the ability of the model to exploit both scores and audio). The main problem is the availability of data for training the models: there is abundance of aligned data, but without the corresponding scores; on the other hand, when scores are available, aligned performances are almost invariably missing. Thus, the choice of the datasets that are included at now has mainly been focused on datasets providing audio, symbolic scores and alignment annotations. However, since datasets fitting all these requirements are quite rare, we wanted to augment the data available to increase the alignment data usable in our research.

\section{Implementation Details}\label{sec:implementation}
\begin{figure*}[t]
    \centering
    \includegraphics[width=0.6\textwidth]{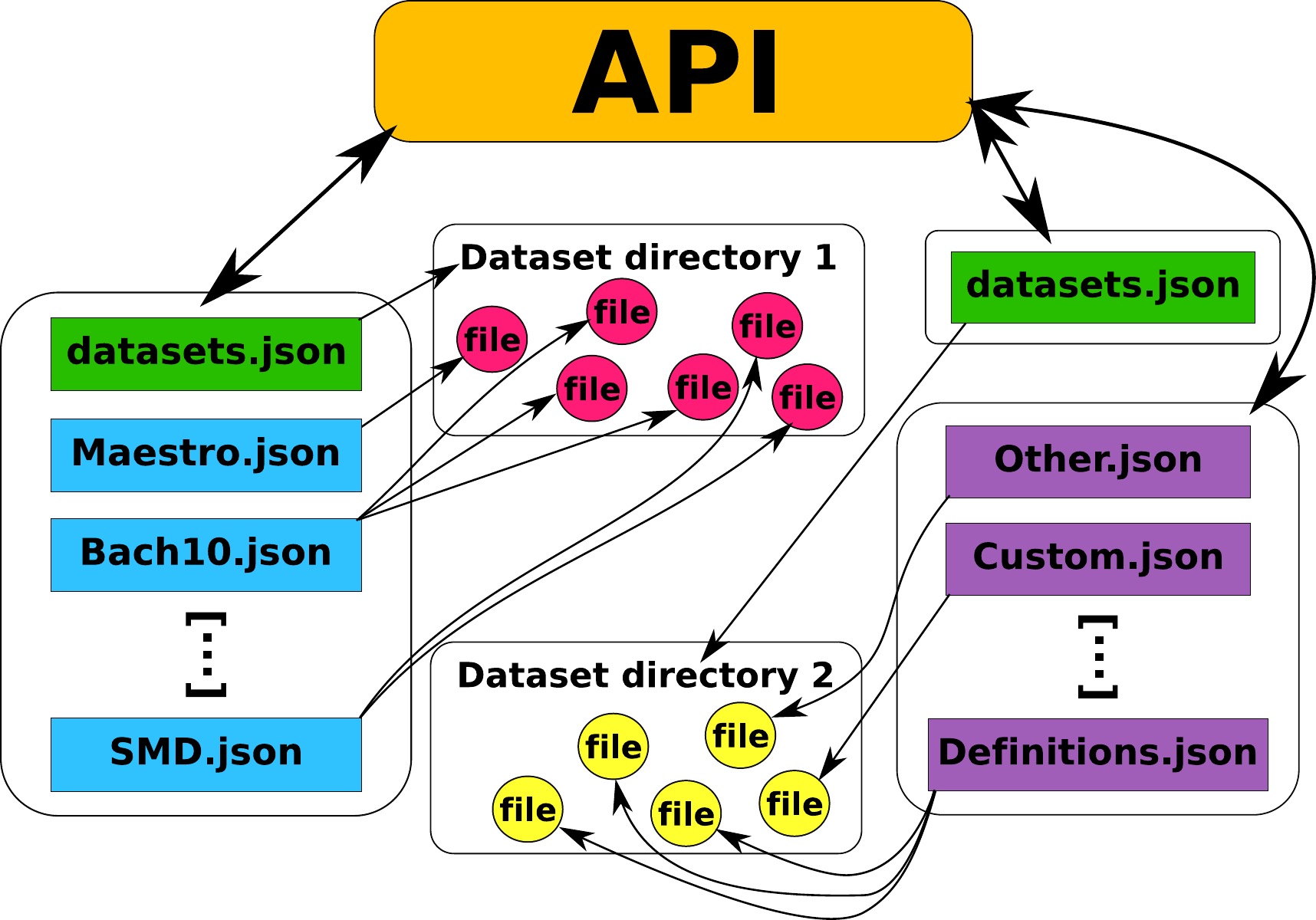}
    \caption{Block diagram of the proposed framework: API interacts with definitions and \texttt{datasets.json}; the former contain references to the actual sound recording files and annotations, while the latter contains references to the dataset root path.}
    \label{fig:framework}
\end{figure*}

This section details the implementation satisfying the design principles outlined in section \ref{sec:design}. Figure \ref{fig:framework} depicts the structure of the overall framework and the interactions between its modules.

\subsection{The datasets.json file}\label{sec:datasets.json}
The entire framework is based on a small-sized but fundamental JSON file loaded by the API and the installation script to get the path where files are installed. Moreover, the user can optionally set a custom directory where to decompress downloaded files if the hard-disk space is a critical issue. Once the installation path is found, the script looks for the existing directories in that path to discover which datasets are already installed and skips them. The API, instead, uses the information of the installation directory to decouple the definition of each single dataset from the directory structure of the user: a user can have the same dataset installed in multiple directories, or use the same dataset from different \textit{datasets.json} without interfering with the API.

\subsection{Definitions}\label{sec:definitions}
In the context of this framework, a \textit{dataset definition} is essentially a JSON\footnote{\url{https://www.json.org/json-en.html}} file which contains generic description of a dataset. \textit{Definitions} are built by using a pre-defined schema allowing the definition of various information useful for the installation of the dataset and for the usage of the API -- e.g. for filtering the dataset. If any of the information is not available for a dataset, the value \texttt{unknown} is offered as well. 

Examples of information contained in \textit{definitions} are:

\begin{compactitem}
    \item \texttt{ensemble}: if the dataset contains solo instrument music pieces or ensemble;
    \item \texttt{instruments}: a list of instruments that are used in the dataset;
    \item \texttt{sources}: if source-separated tracks are available, their format can be added here;
    \item \texttt{recording}: the format of audio recordings;
    \item \texttt{install}: field containing all information for installing the dataset: URL for downloading, shell commands for post-processing data, and so on;
    \item \texttt{ground-truth}: field associated to each type of ground-truth supported by the framework indicating whether the specific  annotation type is available or not -- see Sec~\ref{sec:annotations};
    \item \texttt{songs}: a list of songs with meta-data such as the composer name and instruments used in these songs and with the list of paths to the audio recordings and to the annotations.
\end{compactitem}

%\end{multicols}
%\begin{multicols}{2}

\begin{listing*}
        \begin{minted}[frame=single,framesep=10pt,breaklines,breakafter=d,fontsize=\small]{python}
import audioscoredataset as asd

d = asd.Dataset()
d.filter(instruments=['piano'], ensemble=False, composer='Mozart', ground_truth=['precise_alignment'])

# get audio and all the annotations
audio_array, sources_array, ground_truth_array = d.get_item(1)

# get only the annotations you want
audio_array = d.get_mix(2)
source_array = d.get_source(2)
ground_truth_list = d.get_gts(2)

# get a MIDI Toolbox-like numpy array
mat = d.get_score(2, score_type=['precise_alignment'])
        
# get a pianoroll numpy array
mat = d.get_pianoroll(2, score_type=['non_aligned'])

# or to process songs in parallel using joblib:
def processing(i, dataset, **args, *kwargs):
    mat = d.get_score(2, score_type=['precise_alignment'])
    # other stuffs here
    pass

d.filter(instruments=['violin']).parallel(processing, n_jobs=-1)
        \end{minted}
        \caption{Example of usage for official \textit{definitions}}
        \label{code:1}

\end{listing*}

Once a dataset has been described in this schema, its definition can be used \textit{out-of-the-box} by simply specifying to the API the path of its folder, possibly containing other dataset definitions. All the paths specified in a \textit{definition} must be relative to the installation directory as described in Sec.~\ref{sec:datasets.json}.

For the sake of generalization, we had to deal with a wide heterogeneity in path management among datasets. For instance, Bach10\cite{duan2011soundprism} provides one different annotation file per each instrument in a song; in such a case we list all the annotation files for each song and leave to the API the task  of reassembling them. PHENICX\cite{miron2016scoreinformed}, instead, only provides source-separated tracks and thus we list all of them to reference the mixed track; again, we leave to the API the task of mixing them. In general, we have kept the following principle: if a list of paths is provided where one would logically expect a single path -- such as in mixed tracks or annotation files -- it is intended that the files in the list should be ``merged'' whatever this means for that specific file-type. For instance, if multiple audio recordings are provided instead of only one, it is assumed that the \textit{mixed} track is derivable by adding (and normalizing) all listed tracks; if multiple annotation files are provided, it is assumed that each annotation file refers to a different instrument.

\subsection{Annotations}\label{sec:annotations}
Annotations are added in a custom JSON compressed format stored in the same directory of the audio track that they refer to. In fact, annotation files can be stored anywhere and their path must be provided in the dataset definition relatively to the installation path defined in \textit{datasets.json}. Moreover, one annotation file must be provided for each instrument of the track; if multiple instruments should refer to the same annotations -- e.g. first and second violins -- the annotation file can be only one, but in the dataset definition file, its path should be repeated once for each instrument referring to it.

Multiple types of annotations are available, but not all of them are provided for all the datasets in the official collection. In the dataset definition, the type of annotations available should be explained. In our implementation, we used 3 different levels to describe ground truth availability and reliability:
%\begin{enumerate}%[\bf 1:, labelindent=10pt, leftmargin=*]
\begin{compactenum}[\bfseries {} 1:]
    \setcounter{enumi}{-1} 
    \item annotation type not available
    \item annotation type available and manually or mechanically annotated: this type of annotation has been added by a domain expert or some mechanical transducer -- e.g. Disklavier.
    \item annotation type available and algorithmically annotated: this type of annotation has been added by exploiting a state-of-art algorithm.
\end{compactenum}
%\end{enumerate}

The types of annotations currently supported are:

\begin{compactenum}
\item \textit{precise alignment}: onsets and offsets times in seconds, pitches, velocities and note names for each note played in the recording, taking into account asynchronies inside chords;
\item \textit{broad alignment}: same as \textit{precise alignment} but the alignment does not consider asynchronies inside chords;
\item \textit{non aligned notes}: same as \textit{precise alignment} but not aligned (see \ref{sec:alignment} for more information);
\item \textit{f0}: the f0 of this instrument for each audio frame in the corresponding track;
\item \textit{beats non aligned}: time instances of beats in the non-aligned data;
\item \textit{instrument}: General Midi program number associated with this instrument, starting from 0, while value 128 indicates a drums kit.
\end{compactenum}

\subsection{Alignment}\label{sec:alignment}
As described in section \ref{sec:alignmentprinciple}, this project originated for music alignment research. One problem is the lack of large datasets containing audio recordings, aligned notes and symbolic non aligned scores.

The approach that we used to overcome this problem is to statistically analyze the available manual annotations and to augment the data by approximating them through the statistical model. To prevent biases, we also replaced the manual annotations with the approximated ones. 

For now, the statistical analysis is simple: we compute the mean and the standard deviation of offsets and onsets for each piece. Then, we store the histogram of the standardized offsets and onsets of each note; we also store histograms of the mean and standard deviation values of each piece. To create new misaligned data, we chose a standardized value for each note accompanied by a mean and a standard deviation for each piece, using the corresponding histograms; with these data, we can compute a non-standardized value for each note. Note that initially the histograms are normalized so that they satisfy certain given constraints. In the distributed code, the standardized values are normalized to $1$ (that is, the maximum value is $1$ second), while standard deviations are normalized to $0.2$.

An additional problem is due to the fact that the time units in the aligned data are seconds, while those in the scores are note lengths -- e.g. breve, semibreve and so on. Usually, one translates a note length to seconds by using BPM; however, in some scores the BPM annotation is unavailable or is not reliable. Hence, during the statistical analysis, we always consider the tempo as $20$ BPM, which is a non-usual BPM, in the attempt of minimizing its overall influence. If we used a usual BPM, such as $60$ or $120$, songs with BPM near to that value would have biased the analysis. Moreover, models trained using the produced alignment annotations are ensured to be BPM-independent. Note that one can still try to derive BPM information by making a BPM estimation over the audio\cite{schreiber2018singlestep,bck2015accurate}, a process which highly depends on the algorithm's precision.

\subsection{API}\label{sec:api}
The framework is complemented with a Python API written in Cython.\footnote{\url{https://cython.org/}} It allows in particular to load various dataset definitions aside of the official ones. The API provides methods to retrieve audio and annotations in various structures, such as a matrix list of notes similar to the one used by \textit{Matlab MIDI Toolbox}\cite{eerola2004matlab} or pianorolls. Thanks to the API, one can also filter the loaded datasets' songs based on the original dataset, active instrument, ensemble or solo instrumentation, composer, available annotation types, etc.

Moreover, since the API basically consists in a class representing a large dataset, it is very easy to extend it in order to use it in conjunction of PyTorch or TensorFlow frameworks for training neural network models. In Sec.~\ref{sec:usecases} we provide an example of the specific functionality.

\subsection{Conversion}\label{sec:conversion}

To give the user the ability to write his/her own definitions without having to edit the framework code, we designed a conversion procedure as follows:
\begin{compactenum}
\item the creator can use already developed conversion tools for the most common file formats (MIDI, sonic visualizer, etc.);
\item the creator can still write an ad-hoc function which converts a file from the original format to the ASMD one; in this case the creator has to decorate the conversion function with a special decorator provided by ASMD;
\item the creator adds the needed conversion function in the \texttt{install} section in the dataset definition;
\item the user can run the conversion script for only a specific dataset or for all other datasets.
\end{compactenum}
All the technical details are available in the official documentation.\footnote{\url{https://asmd.readthedocs.io/}}

    \begin{listing*}[t]
        \begin{minted}[frame=single,framesep=10pt,breaklines,breakafter=d,fontsize=\small]{python}
import audioscoredataset as asd

d = asd.Dataset(['path/to/directory/containing/custom/definitions', 'path/to/the/official/definitions/'])
d.filter(instruments=['piano'], ensemble=False, composer='Mozart', ground_truth=['precise_alignment'])
    
        \end{minted}
        \caption{Example of usage for custom \textit{definitions}}
        \label{code:2}
    \end{listing*}
    
    \begin{listing*}[t]
        \begin{minted}[frame=single,framesep=10pt,breaklines,breakafter=d,fontsize=\small]{python}
import torch
import audioscoredataset as asd
from torch.utils.data import Dataset as TorchDataset

class MyDataset(asd.Dataset, TorchDataset):
    def __init__(self, *args, **kargs):
        super().__init__(['path/to/definitions']).filter(instruments=['piano'])
        
    def __getitem__(self, i):
        # for instance, return the MIDI Toolbox-like score
        return torch.tensor(self.get_score(i))
        
    def another_awsome_method(self, *args, **kargs):
        print("Hello, world!")
        
for i, mat in enumerate(MyDataset()):
    # train your nn model here
    
        \end{minted}
        \caption{Example for using ASMD inside PyTorch}
        \label{code:3}
    \end{listing*}

\section{Use Cases}\label{sec:usecases}
This section demonstrates the efficacy of the ASMD framework through four different use cases.

\subsection{Using API with the official dataset collection}

To use the API, the user should carry out the following steps:

\begin{compactitem}
    \item import \texttt{audioscoredataset};
    \item create a \texttt{audioscoredataset.Dataset} object, giving the path of the \texttt{datasets.json} file as an argument to the constructor;
    \item use the \texttt{filter} method on the object to filter data according to his/her needs (conveniently, it is also possible to re-filter them at a  later stage, without reloading the \texttt{datasets.json} file);
    \item retrieve elements by calling the \texttt{get\_item} method or similar ones.
\end{compactitem}

After the execution of the \texttt{filter} method, the \texttt{Dataset} instance will contain a field \texttt{paths} representing the list of correct paths to the files requested by the user. Listing \ref{code:1} shows an example of such an operation.

\subsection{Using API with definitions for a customized dataset}

Whenever the user wishes to apply customized definitions, he/she need simply to provide the list of directories to the \texttt{Dataset} constructor, as shown in listing \ref{code:2}.
%\end{multicols}

    \begin{listing*}[t]
    
        \begin{minted}[frame=single,framesep=10pt,breaklines,breakafter=d,fontsize=\small]{python}
from audioscoredataset.convert_from_file import convert, prototype_gt
from copy import deepcopy

# use @convert
@convert(['.myext'])
def function_which_converts(filename, *args, **kargs):
    # prepare empty output
    out = deepcopy(prototype_gt)
    
    # open file
    data = csv.reader(open(filename), delimiter=',')
    
    # fill output dictionary
    for row in data:
        out[alignment]["onsets"].append(float(row[0]))
        out[alignment]["offsets"].append(float(row[0]) + float(row[2]))
        out[alignment]["pitches"].append(int(row[1])
    
    return out
        \end{minted}
    \caption{Example for writing a custom conversion function}
    \label{code:4}
    \end{listing*}

%\begin{multicols}{2}

\subsection{Using ASMD with PyTorch}

Integrate \textit{ASMD} with \textit{PyTorch} is straightforward. The user has to inherit from both PyTorch and ASMD \texttt{Dataset} classes and to implement the \texttt{\_\_getitem\_\_} method. Listing \ref{code:3} shows such an example.

\subsection{Writing a conversion function and a custom dataset definition}
Towards adding new definitions enabling users to download datasets, a user should also provide a conversion function. Listing \ref{code:4} is an example of one can write its own conversion function. However, conversion functions for the most common file types -- such as Midi and Sonic Visualizer -- are already provided by the framework.

\section{Conclusions}\label{sec:conclusions}
Future works will focus on the enhancement of conversion and installation procedures, as well as on the definition of standards for music annotations. In addition, multimodal music processing often requires processing of annotation types not included in this version of the framework, but could instead be handled in a future release. Some annotation types could be stored in standalone formats and users should be able to distribute annotations focusing only on a specific ground truth kind, thus enhancing the distributed infrastructure of ASMD.

Studying the user experience of the framework should also be a priority: for instance, users could be able to choose datasets also based on the estimation of the download time since for some datasets that is a big issue. Labels used in the annotation format are also relevant to ease the usage of the framework by new users, especially in a multidisciplinary field such as the sound and music computing.

This paper presented a new framework for multimodal music processing. We hope that our efforts in easing the development of multimodal machine learning approaches for music information processing will be useful to the sound and music computing community. We are completely aware that for a truly general and usable framework, the participation of various and different perspectives is needed and we are therefore open to any contribution towards the creation of utilities that allow training and testing multimodal models, ensuring reasonable generalization ability and reliable reproducibility of scientific results.

\begin{acknowledgments}
We would like to thank all people that worked on the datasets used in the ASMD framework: Bach10\cite{duan2011soundprism}, Maestro\cite{hawthorne2019enabling}, MusicNet\cite{thickstun2018invariances}, PHENICX Anechoic\cite{miron2016scoreinformed}, SMD\cite{muller2011saarland}, Traditional Flute Dataset\cite{brum2018traditional}, TRIOS\cite{fritsch2012trios} and Vienna 4x22 Piano Corpus\cite{goebl1999vienna}.
\end{acknowledgments} 

%%%%%%%%%%%%%%%%%%%%%%%%%%%%%%%%%%%%%%%%%%%%%%%%%%%%%%%%%%%%%%%%%%%%%%%%%%%%%
%bibliography here
\bibliography{smc2020bib}
\balance
%\end{multicols}
\end{document}